\documentstyle[aps,graphicx]{revtex}

\begin{document}
\preprint{\vbox{\hbox{\bf UCD-98-16} }}
\title{A Monte Carlo Study of the Rare Decay $B \to X_s \ell^+ \ell^-$}
\author{Kingman Cheung}
\address{
Department of Physics, University of California at Davis, Davis, CA 95616}
\date{\today}
\maketitle

\begin{abstract} 
We study, using a monte carlo approach, the rare decay $B\to X_s
\ell^+ \ell^-$ including effects of the arbitrariness of the phase
between the $\psi$ amplitudes and the perturbative amplitude,
$b$-quark fermi motion inside the $B$ meson, and experimental
smearing of lepton momenta.  The fermi motion of $b$-quark inside the
$B$ meson is modeled by the ACCMM model.  We found that such effects
reduce the sensitivities of the spectra of invariant mass
and forward-backward asymmetry of the lepton pair to new
physics; especially, in the neighborhood of the $\psi$ resonances.
We also estimate the sensitivity range of Wilson coefficients with respect
to the uncertainties.
\end{abstract}

\pacs{PACS numbers: }

\section{Introduction} 

Rare $B$-meson decays are of immense interests, especially after the
first discovery of the penguin decay $B\to X_s \gamma$ by CLEO \cite{cleo}.
No other rare $B$ decays have been identified yet, but they are about at
the door of discovery.
One of these decays that involves a lepton pair in the final state 
$B\to X_s \ell^+ \ell^-\; (\ell=e,\mu)$ is especially interesting. 
It is potentially a window to new physics beyond the standard model (SM)
and has been studied extensively in literature.
In particular, the spectra of invariant mass and forward-backward 
asymmetry of the lepton pair are shown very sensitive to different
types of new physics \cite{back}.
QCD corrections have been calculated and are under control,
which enables one to predict reliably the decay rate and spectra \cite{buras}.
The calculation including QCD corrections is done by 
an effective hamiltonian approach, which we shall describe briefly in the next
section.  However, a complication arises from a long distance contribution
of $B\to X_s J/\psi \to X_s \ell^+ \ell^-$.  
To include this contribution the amplitudes of $\psi$ resonances are added 
to the perturbative amplitude in a rather phenomenological way \cite{psi}.
Each of the $\psi$ amplitudes has an overall normalization to be determined
by experiments, an arbitrary phase relative to the perturbative amplitude,
\footnote{Although an argument \cite{donnell}
of unitarity indicates that the phase should
be unity, we shall, however, allow a more general phase in order to fully
estimate this uncertainty.}
and the $\psi$ resonance shape is described by a Breit-Wigner prescription.
All these give rise to uncertainties in the prediction of the spectra.

Another source of uncertainty to decay spectra comes from the fact 
that $B$ meson is a bound state of a $b$ quark and a light quark. 
The bound state effect can be represented by a fermi motion of the
$b$ quark, which is of order of $\Lambda_{\rm QCD}$.  
We use the popular model of Altarelli {\it et al.} \cite{altarelli} 
to formulate the fermi motion.  Another important 
smearing effect comes from the resolution in the measurement of 
lepton momenta.  

The objective of this report is to investigate the effects of 
(i) the arbitrariness of the phase between the $\psi$ resonance amplitudes
and the perturbative amplitude, 
(ii) fermi motion of the $b$ quark inside the $B$ meson, and 
(iii) the smearing of lepton momenta on the 
spectra of invariant mass and forward-backward asymmetry of the lepton pair.  
We show that the sensitivity is
reduced, in particular, in the regions next to $\psi$ resonances.  
Very often new physics can be parameterized by Wilson coefficients $C_i(M_W)$.
We shall investigate how much the $C_i(M_W)$ are needed to change 
such that the spectra are distinguishable from the SM ones, under the
effects of the above.  

The organization of the paper is as follows.  In the next section, we
describe briefly the calculation framework.  In sec. III, we shall describe
the smearing effects of fermi motion of $b$ quark and lepton momentum
measurements.  In sec. IV, we estimate the
sensitivities of the leptonic spectra to variation of $C_i(M_W)$.  We
conclude in sec. V.

\section{Calculation Framework}

The detail description of the effective hamiltonian approach can be found in
Refs. \cite{wise,buras}.  Here we present the highlights that are relevant
to our discussions.  The effective hamiltonian for $B\to X_s\ell^+ \ell^-$ 
at a factorization scale of order $O(m_b)$ is given by \cite{buras,wise,ali}
\begin{equation}
\label{eff}
{\cal H}_{\rm eff} = - \frac{4 G_F}{\sqrt{2}} V^*_{ts} V_{tb} \biggr [
\sum_{i=1}^6 \; C_i(\mu) O_i(\mu) + 
C_{7\gamma}(\mu) O_{7\gamma}(\mu) +
C_{8G}(\mu) O_{8G}(\mu) +
C_9 (\mu) O_9(\mu) +
C_{10} (\mu) O_{10}(\mu) 
\biggr ] \;.
\end{equation}
The operators $O_i$ can be found in Ref. \cite{buras,ali}, 
of which the $O_1$ and $O_2$ are the current-current operators and 
$O_3-O_6$ are the QCD penguin
operators.  $O_{7\gamma}$ and $O_{8G}$ are, respectively, the magnetic
penguin operators specific for $b\to s\gamma$ and $b\to s g$.  
Since $O_9$ and $O_{10}$ are directly involved in the decay 
$b \to s \ell^+ \ell^-$, we list them here:
\begin{equation}
O_9 = \frac{e^2}{16 \pi^2} \bar s_\alpha \gamma^\mu L b_\alpha \; 
\bar \ell \gamma_\mu \ell \;, 
\qquad 
O_{10} = \frac{e^2}{16 \pi^2} \bar s_\alpha \gamma^\mu L b_\alpha \; 
\bar \ell \gamma_\mu \gamma_5 \ell \;,
\end{equation}
where $L(R) = (1\mp \gamma_5)/2$. 
Here we neglect the mass of the external strange quark compared to the external
bottom-quark mass.  

The factorization in Eq.(\ref{eff}) facilitates the separation of the 
short-distance and long-distance parts, of which the short-distance parts
correspond to the Wilson coefficients $C_i$ and are calculable
by perturbation while the long-distance parts correspond to the operator
matrix elements.  
The physical quantities based on Eq.~(\ref{eff}) should
be independent of the factorization scale $\mu$.  The natural scale for 
factorization is of order $m_b$ for the decay $B\to X_s \ell^+ \ell^-$.
First, at the electroweak scale, say $M_W$, the full theory is matched onto
the effective theory and the coefficients $C_i(M_W)$ at the $W$-mass scale
are extracted in the matching process.  
Second, the coefficients $C_i(M_W)$ at the $W$-mass scale are evolved 
down to the bottom-mass scale using renormalization group equations.  
Since the operators $O_i$'s are all mixed under renormalization, the 
renormalization group equations for $C_i$'s are a set of coupled 
equations:
\begin{equation}
\vec C(\mu) = U(\mu,M_W) \vec C(M_W) \;,
\end{equation}
where $U(\mu,M_W)$ is the evolution matrix and $\vec C(\mu)$ is the vector 
consisting of $C_i(\mu)$'s.  The calculation of the entries of the evolution
matrix $U$ is nontrivial but it has been written down completely in the
leading order \cite{buras}.  The coefficients $C_i(\mu)$ at the scale $O(m_b)$
are given by \cite{buras}
\begin{eqnarray}
C_j(\mu) &=& \sum_{i=1}^8 k_{ji} \eta^{a_i} \qquad (j=1, ..., 6) \;\;,\\
C^{\rm eff}_{7\gamma}(\mu) &=& 
\eta^{\frac{16}{23}} C_{7\gamma}(M_W) + \hbox{$8\over3$}
\left(\eta^{\frac{14}{23}} - \eta^{\frac{16}{23}} \right ) C_{8G}(M_W) 
+ C_2(M_W) \sum_{i=1}^8 h_i \eta^{a_i} \;\;,\\
C^{\rm eff}_{8G} (\mu) &=& 
\eta^{\frac{14}{23}} C_{8G}(M_W) + C_2(M_W) \sum_{i=1}^8 
\bar h_i \eta^{a_i} \;\;, \\
C^{\rm eff}_9 (\mu, \hat s)&=& 
C_9(\mu) \eta(\hat s) + Y(\hat s) + Y_{\rm res}(\hat s) \;, \\
C_{10} (\mu) &=& C_{10}(M_W) \;,
\end{eqnarray}
with $\eta=\alpha_s(M_W)/\alpha_s(\mu)$, $\hat s = q^2/m_b^2$,
and $q^2$ is the invariant mass squared of the lepton pair.
The $a_i$'s, $k_{ji}$'s,
$h_i$'s, and $\bar h_i$'s can be found in Ref.~\cite{buras}.  
The functions $C_i(M_W)\, (i=1-10), C_9(\mu), \eta(\hat s)$, and 
$Y(\hat s)$ can be found in Refs. \cite{buras,ali}.
Here we pay more attention to the term $Y_{\rm res}(\hat s)$, which is the 
contribution from the $c\bar c$ resonances:
\begin{equation}
\label{psi}
Y_{\rm res}(\hat s) = - K \frac{3\pi}{\alpha^2} C^{(0)} \,
\sum_{i=\psi(1S),...,\psi(6S)} 
\frac{\Gamma( \psi_i \to \ell^+ \ell^-) M_{\psi_i}}
{\hat s m_b^2 - M_{\psi_i}^2 + i M_{\psi_i} \Gamma_{\psi_i} } \;,
\end{equation}
where $C^{(0)}\equiv 3C_1+C_2+3C_3+C_4+3C_5+C_6$ and 
the parameter $K$ is set at $|K|=2.3$ \cite{ligeti,ali} with a phase 
$\kappa$, and we vary the phase 
$\kappa$ to allow for the uncertainty in adding this long-distance 
contribution to the perturbative amplitude.  
In subsequent discussions, we only include the $\psi(1S)$ and $\psi(2S)$
resonances in our calculation for simplicity.  In Eq. (\ref{psi}), the
resonance shape is described by a scale-independent Breit-Wagner prescription.
We have verified that if the width $\Gamma_\psi$
in Eq. (\ref{psi}) is replaced by a $q^2$-dependent width 
$\Gamma_\psi(q^2) =\Gamma_\psi (q^2/M_\psi^2)$ there are no visible changes
to our results, because the width is very narrow.

With the hamiltonian we can write down the decay amplitude for
$b \to s \ell^- \ell^+$ and the spin-averaged square of the amplitude is 
given by
\begin{eqnarray}
\overline{\sum} |A|^2 &=& \frac{4 G_F^2 \alpha^2}{\pi^2}\, |V_{ts}^* V_{tb}|^2 
\nonumber \\
&\times & \biggr[
\left( |C_9^{\rm eff} -C_{10}|^2 -4 \frac{|C_7^{\rm eff}|^2}{\hat s} \right )
 b\cdot \ell^+ 
s\cdot \ell^-
+ \left( |C_9^{\rm eff} +C_{10}|^2 -4 \frac{|C_7^{\rm eff}|^2}{\hat s} \right )
b\cdot \ell^-
s\cdot \ell^+ \nonumber \\
&&+ \{ \frac{2}{\hat s} \left( C_7^{\rm eff} ( {C_9^{\rm eff}}^* - C^*_{10}) 
+ {C_7^{\rm eff}}^*(C_9^{\rm eff} - C_{10})
\right ) + 8 |C_7^{\rm eff}|^2  \frac{b\cdot q }{{\hat s}^2 m_b^2} \}
\ell^- \cdot \ell^+ \, s\cdot \ell^-  \nonumber \\
&&+ \{ \frac{2}{\hat s} \left( C_7^{\rm eff} 
( {C_9^{\rm eff}}^* + C^*_{10}) + {C_7^{\rm eff}}^*(C_9^{\rm eff} + C_{10})
\right ) + 8 |C_7^{\rm eff}|^2  \frac{b\cdot q }{{\hat s}^2 m_b^2} \}
\ell^- \cdot \ell^+ \, s\cdot \ell^+  
\biggr ] \;,
\end{eqnarray}
where the momenta of the particles are labelled by the particle symbols
and $q=\ell^+ + \ell^-$.  
Since the decay width depends on the fifth power of $m_b$, a small
uncertainty in $m_b$ will create a large uncertainty in the decay rate, 
therefore, the decay rate is often normalized by the experimental
semi-leptonic width:
\begin{equation}
\label{width}
\frac{1}{ \Gamma(b \to X_c \ell \nu)}
\frac{d \Gamma (b \to s \ell^+ \ell^-)}{d \hat s}
= \frac{\alpha^2}{4 \pi^2} \left| \frac{V_{ts}^* V_{tb}}{V_{cb}}
\right |^2 \, \frac{(1-\hat s)^2}{f(m_c/m_b)} \,
\biggr[ \left( |C_9^{\rm eff}|^2 + |C_{10}|^2 \right ) (1+2\hat s) +
4|C_7^{\rm eff}|^2 \frac{2+\hat s}{\hat s} 
+ 12 {\cal R}\left ( C_7^{\rm eff} {C_9^{\rm eff}}^* \right )
\biggr ] \;,
\end{equation}
where $f(z)= 1-8z^2 +8z^6 -z^8 -24z^4 \ln z$.
The lepton forward-backward asymmetry $A$ is defined by the angle $\theta$ 
between the $\ell^-$ and the $b$-quark in the center-of-mass frame of 
the lepton pair:
\begin{equation}
\label{asy}
\frac{dA}{d \hat s} \equiv 
 \frac{ \int_0^1 d\cos\theta \frac{d\Gamma}{d\hat s} -
        \int_{-1}^0 d\cos\theta \frac{d\Gamma}{d\hat s} }
{ \int_0^1 d\cos\theta \frac{d\Gamma}{d\hat s} +
  \int_{-1}^0 d\cos\theta \frac{d\Gamma}{d\hat s} }
= \frac{ 3[\hat s {\cal R}(C^{\rm eff}_9 C_{10}^*) + 2 {\cal R}(C_7^{\rm eff}
C_{10}^* )] }
{ \left( |C_9^{\rm eff}|^2 + |C_{10}|^2 \right ) (1+2\hat s) +
4|C_7^{\rm eff}|^2 \frac{2+\hat s}{\hat s} 
+ 12 {\cal R}\left ( C_7^{\rm eff} {C_9^{\rm eff}}^* \right ) } \;.
\end{equation}

One comment on these spectra under Lorentz transformation is given as follows.
Since in the
following we shall include the fermi motion of the $b$ quark, we have to
boost the spectra back to the $B$ meson rest frame.  In addition, the 
$B$ meson is not at rest in the laboratory frame, the spectra may as well be
boosted to the laboratory frame.  
Since the quantity $\hat s$ in Eqs. (\ref{width}) and (\ref{asy}) is 
Lorentz-invariant, the spectrum in Eq. (\ref{width}) is almost
stable under Lorentz boost.  The slight change is due to the fact that
different fermi-motion momentum $p$ will give different $m_b$, which
in turns affects the invariant mass spectrum.  
On the other hand, the forward-backward asymmetry is defined solely in
the rest frame of the lepton pair because the angle between the lepton
$\ell^-$ and the $b$ quark is not Lorentz-invariant.  

\section{Smearing}

The free quark model that treats the decay of a $B$ meson as a free $b$ quark
is an idealistic model in the sense that it is only 
correct if $m_b$ is infinitely
heavy.  It has been proved by using heavy quark effective theory that
the correction is of order $1/m_b^n$; in particular the lepton spectrum
starts with $n=2$ \cite{hq}.  This behavior can also be understood in terms
of fermi motion (FM) of the $b$ quark inside the $B$ meson, which 
causes a small offshellness of the $b$ quark.  The FM model, often called
ACCMM model \cite{altarelli}, is characterized by two parameters $p_F$ and
the spectator quark mass $m_{\rm sp}$.  The $b$ quark is assumed to have a
small momentum $p$, which follows a gaussian distribution with the
parameter $p_F$:
\begin{equation}
\phi(p) = \frac{4}{\sqrt{\pi} p_F^3}\, \exp\left( \frac{-p^2}{p_F^2} \right)\;,
\end{equation}
and a normalization $\int_0^\infty dp\; p^2 \phi(p) =1$.
Energy-momentum conservation requires the $b$ quark mass to be dependent on
$p$:
\begin{equation}
\label{mb}
m_b^2 (p) = m_B^2 + m_{\rm sp}^2 - 2m_B \sqrt{p^2 + m_{\rm sp}^2} \;.
\end{equation}
As a consequence of this continuous range of $m_b$, spectra will be
smeared.  However, the invariant mass spectrum of the lepton pair will be
affected minimally because the invariant mass is a Lorentz-invariant 
quantity and the lepton spectrum only receives corrections of order $1/m_b^2$.
The invariant mass spectrum with the effect of FM smearing is given by
\begin{equation}
d \Gamma(B \to X_s \ell^+ \ell^-) =
\int_0^{p_{\rm max}} dp \; p^2 \phi(p) \; 
\frac{d \Gamma}{d\hat s} \, \frac{dq^2}{m_b^2} \;,
\end{equation}
and a similar expression is valid for the lepton forward-backward asymmetry.
The parameter $p_{\rm max}$ is chosen such that the minimum for $m_b(p)$ in
Eq. (\ref{mb}) goes to $M_\psi$.
In our numerical calculation, we use a set of values for $p_F$ and 
$m_{\rm sp}$, which is consistent with the
results obtained in the spectra of $B\to X_s\gamma$, $B\to X \ell \nu$, and 
$B\to X_s \psi$ \cite{ali}:
\begin{equation}
p_F = 0.54 \; {\rm GeV} \qquad {\rm and} \qquad  m_{\rm sp} =0.15\; {\rm GeV}
\;.
\end{equation}

While the FM smearing is of theoretical in nature, 
another smearing effect comes from the measurement of lepton
energies and momenta.  This smearing effect is actually stronger than the
FM smearing.  Note that both the angular and energy measurements will be
affected by detector resolution.  We shall employ the following resolutions, 
which are used in the CLEO $B\to X_s \psi$ measurement \cite{cleo-psi}, in our
study:
\begin{equation}
\frac{\delta E}{E} = \left( \frac{0.35}{E^{0.75}} + 1.9 -0.1E \right )
 \% \;, \qquad
\left(\frac{\delta p_t}{p_t}\right)^2 = (0.0015 p_t)^2 + 0.005^2 \;,
\end{equation}
where $E$ and $p_t$ are in GeV. 
In our study, the $b$-quark momentum $p$ and its direction are chosen
randomly and the above resolutions are imposed on the final state lepton
momenta in a event-by-event basis.  The advantage of this monte carlo 
approach is that both FM smearing and lepton momentum smearing can be
combined in a event-by-event basis.

\begin{figure}[th]
\leavevmode
\begin{center}
\includegraphics[width=3.5in]{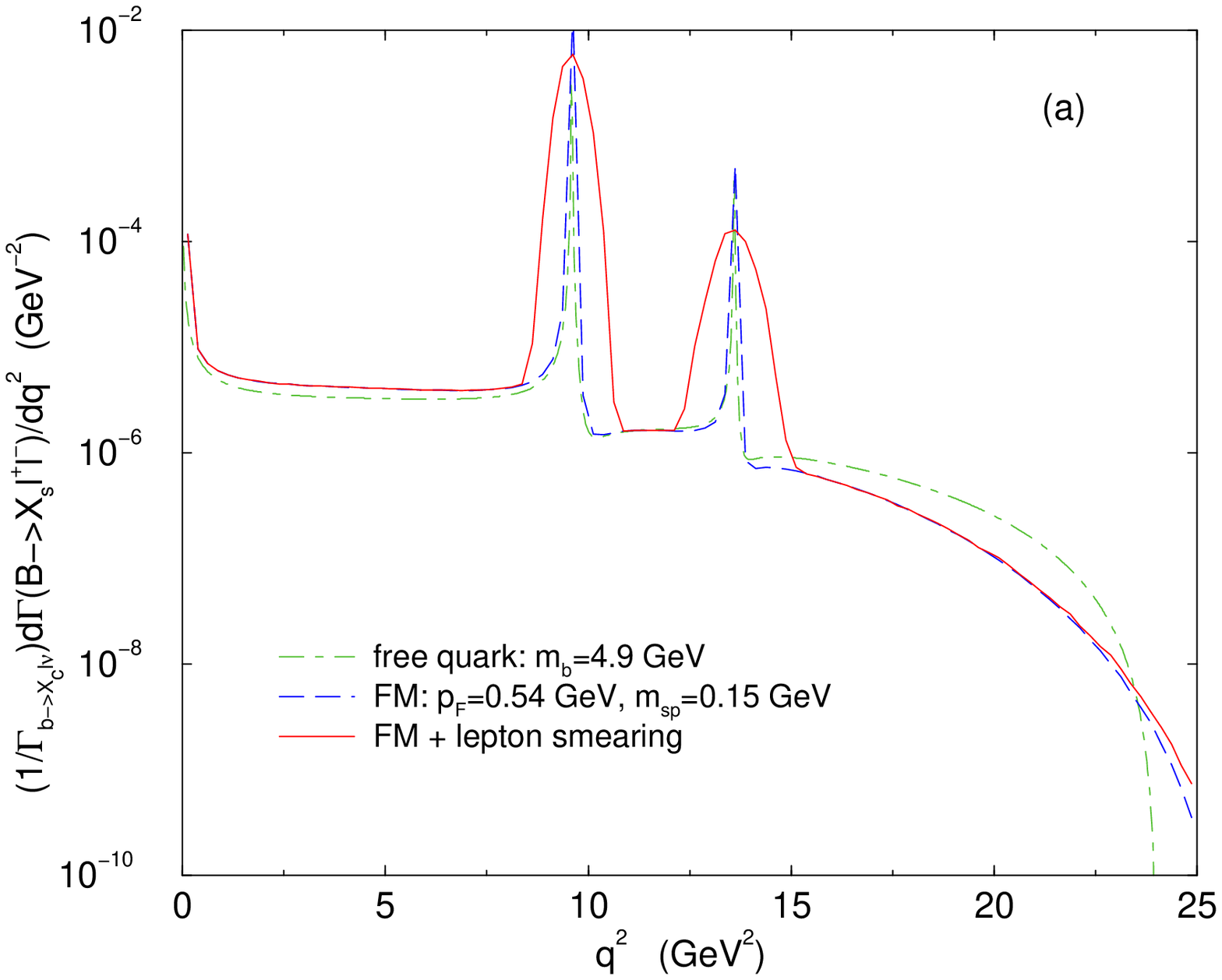}
\includegraphics[width=3.5in]{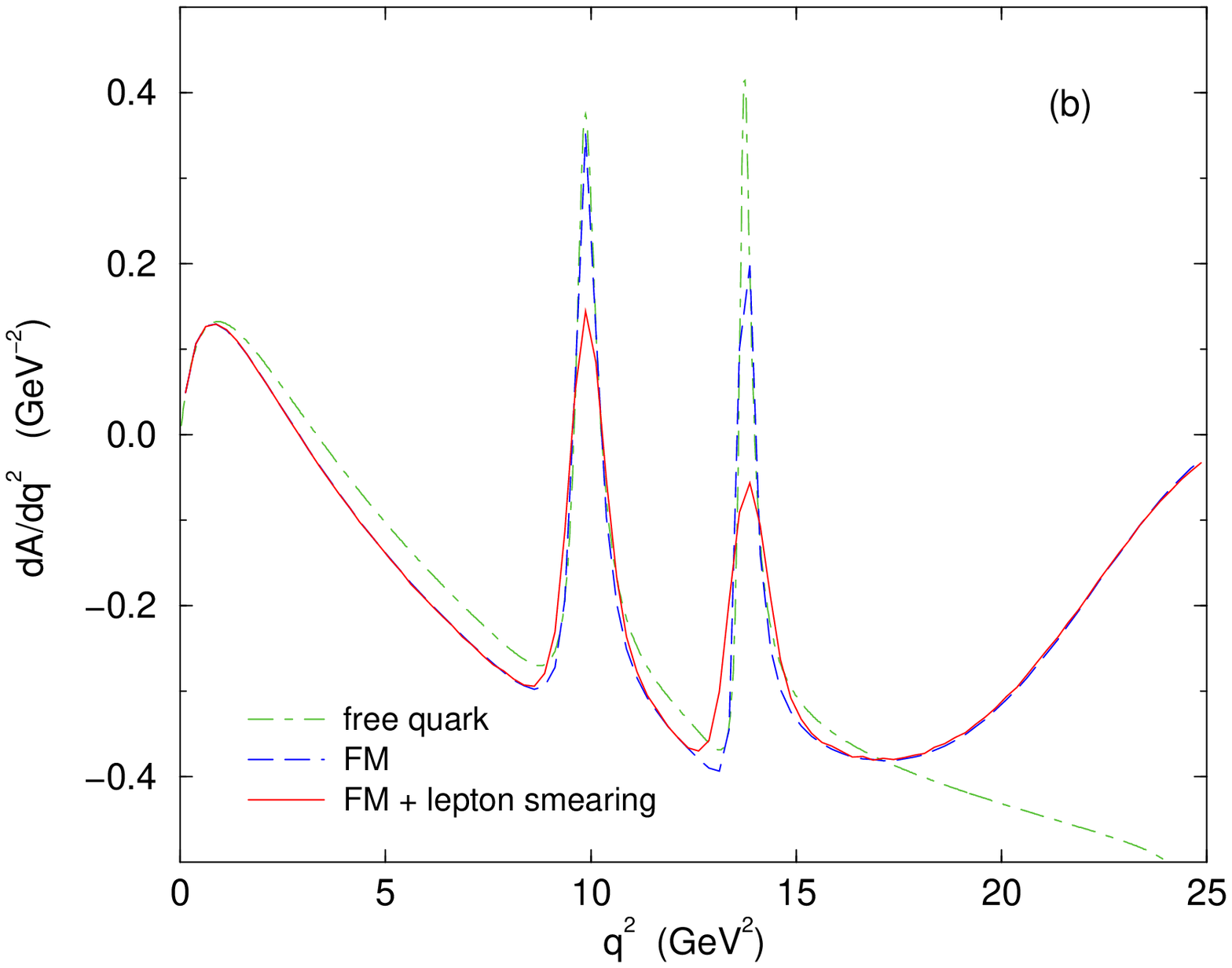}
\end{center}
\caption{ \label{fig1}
The distributions (a) $(1/\Gamma_{b\to X_c \ell \nu})
d\Gamma(B\to X_s \ell^+ \ell^-)/dq^2$ and (b) $dA/dq^2$ showing the effects
of fermi motion (FM) of the $b$ quark inside $B$ meson and of the leptonic
smearing.
}
\end{figure}

\begin{figure}[th]
\leavevmode
\begin{center}
\includegraphics[width=3.5in]{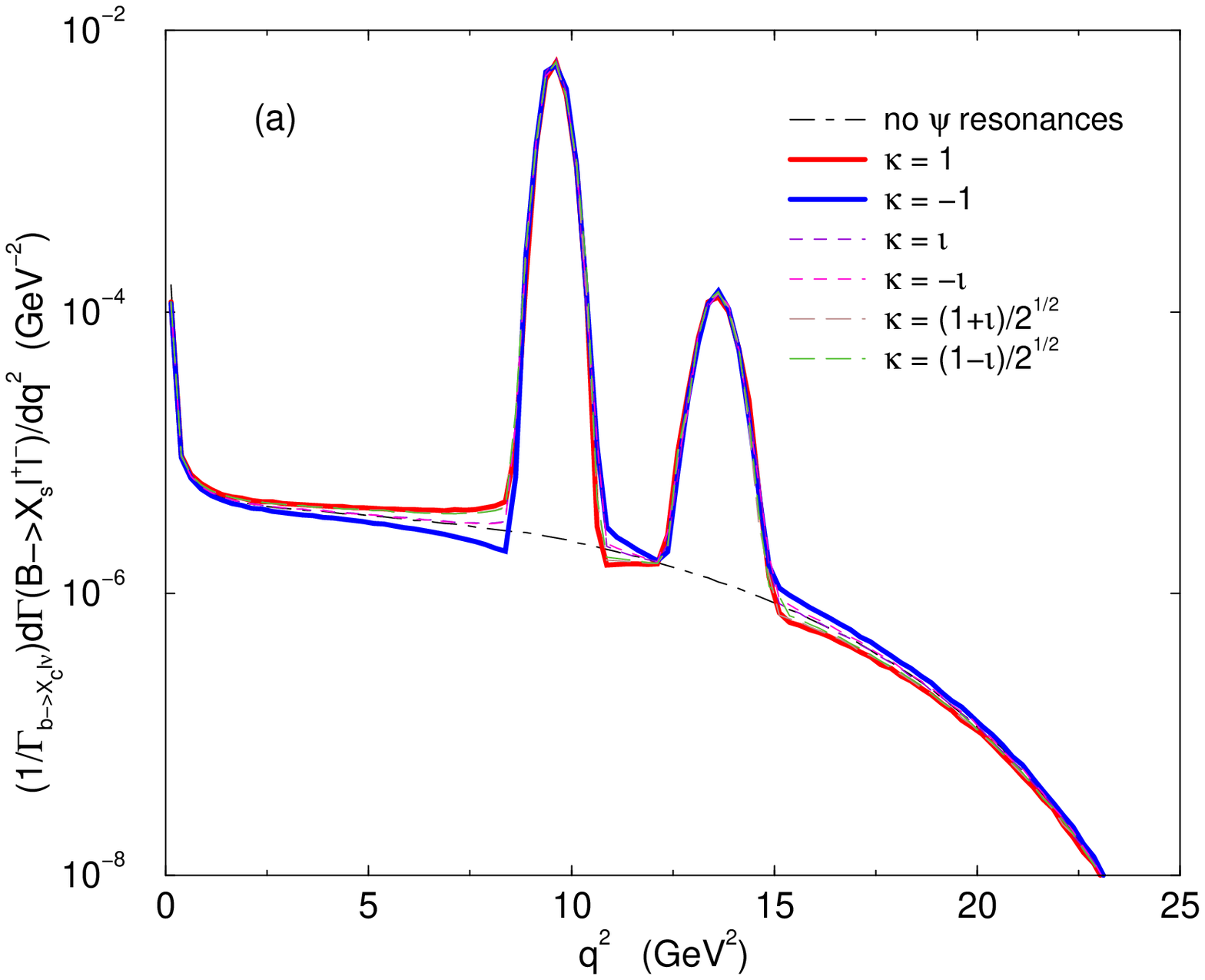}
\includegraphics[width=3.5in]{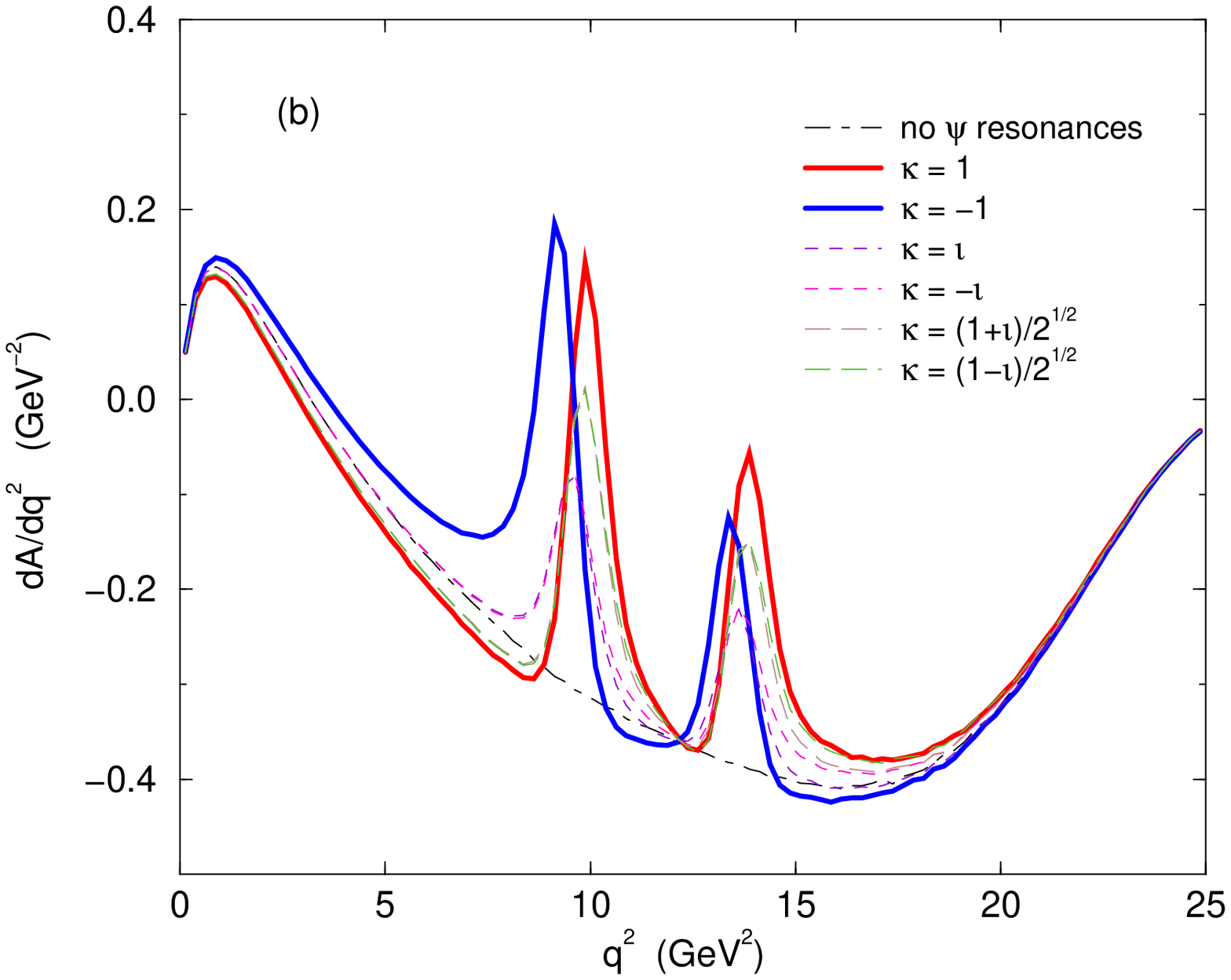}
\end{center}
\caption{
\label{fig2}
The distributions (a) $(1/\Gamma_{b\to X_c \ell \nu})
d\Gamma(B\to X_s \ell^+ \ell^-)/dq^2$ and (b) $dA/dq^2$ showing the effects
of varying the phase $\kappa$ between the $\psi$ and perturbative amplitudes.
}
\end{figure}

The smearing effects of FM and lepton momentum measurements 
are demonstrated in Figs. \ref{fig1}(a) and (b).  
It is clear that the regions around the
resonance peaks are smeared quite significantly.
In Figs. \ref{fig2}(a) and (b), we show the effect of varying the phase
$\kappa$ between the $\psi$ amplitudes and the perturbative amplitude.
We show the results for $\kappa=\pm1, \pm i, (1\pm i)/\sqrt{2}$.  For
simplicity we treat the phases for $\psi(1S)$ and $\psi(2S)$ to be the 
same.  One can see
that $\kappa=\pm 1$ allows for maximal interference with the perturbative
amplitude.  Anywhere in between is possible.  We then treat the region
roughly bounded by $\kappa=\pm 1$ curves as the uncertainty in prediction
in our following discussions.

\section{Sensitivity to New Physics}

We use the two-Higgs-doublet-model II (2HDMII) 
and a model-independent method 
to illustrate the sensitivities.  We start with the 2HDMII, which
is a popular extension of the SM and provides a minimal Higgs sector
for supersymmetry.  The extra
contributions to the Wilson coefficients $C_i$ depend on the charged Higgs
boson mass and $\tan\beta$.  For clarity we list the coefficients
$C_{7-10}(M_W)$ here \cite{buras,ali,hewett}:
\begin{eqnarray}
C_7(M_W) &=& - \frac{A(x_t)}{2} - \frac{A(x_H)}{6\tan^2\beta} - B(x_H) \\
C_8(M_W) &=& - \frac{D(x_t)}{2} - \frac{D(x_H)}{6\tan^2\beta} - E(x_H) \\
C_9(M_W) &=& \frac{Y}{\sin^2\theta_{\rm w}} - 4 Z \\
C_{10}(M_W) &=& -\frac{Y}{\sin^2\theta_{\rm w}} \\
Y &=& C(x_t) - F(x_t) - \frac{x_t}{8 \tan^2\beta} f_5(x_H) \\
Z &=& C(x_t) + \frac{G(x_t)}{4} - \frac{x_t}{8 \tan^2\beta} f_5(x_H) 
             - \frac{1}{72 \tan^2\beta} f_6(x_H)  \;,
\end{eqnarray}
where
\begin{eqnarray}
A(x) &=& x \biggr[ \frac{8 x^2 + 5 x -7}{12(x-1)^3} - 
             \frac{(3x^2 - 2x)\log x}{2(x-1)^4} \biggr] \\ 
B(x) &=& x \biggr[ \frac{5x -3}{12(x-1)^2} - 
              \frac{(3x -2)\log x}{6(x-1)^3} \biggr] \\
D(x) &=& x \biggr[ \frac{x^2 -5x -2}{4(x-1)^3} +
                 \frac{3x \log x}{2(x-1)^4} \biggr ] \\
E(x) &=& x \biggr[ \frac{x-3}{4(x-1)^2} + \frac{\log x}{2(x-1)^3} \biggr ]\\
C(x) &=& \frac{x}{8} \biggr[ \frac{x-6}{x-1} + 
                    \frac{(3x +2)\log x}{(x-1)^2}  \biggr ] \\
F(x) &=& \frac{1}{4} \biggr[\frac{x}{1-x} + \frac{x\log x}{(x-1)^2} \biggr]\\
G(x) &=& -\frac{4}{9} \log x  + \frac{-19x^3 + 25x^2}{36(x-1)^3} 
       + \frac{x^2(5x^2 -2x -6)\log x}{18(x-1)^4} \\
f_5(x) &=& \frac{x}{1-x} + \frac{x \log x}{(1-x)^2} \\
f_6(x) &=& \frac{38x -79x^2 +47x^3}{6(1-x)^3} +
             \frac{(4x -6x^2 +3x^4)\log x}{(1-x)^4} \\
x_t    &=& m_t^2 / M_W^2 \;, x_H= m_t^2 / m_{H^\pm}^2 \;.
\end{eqnarray}

\begin{figure}[th]
\leavevmode
\begin{center}
\includegraphics[width=3.5in]{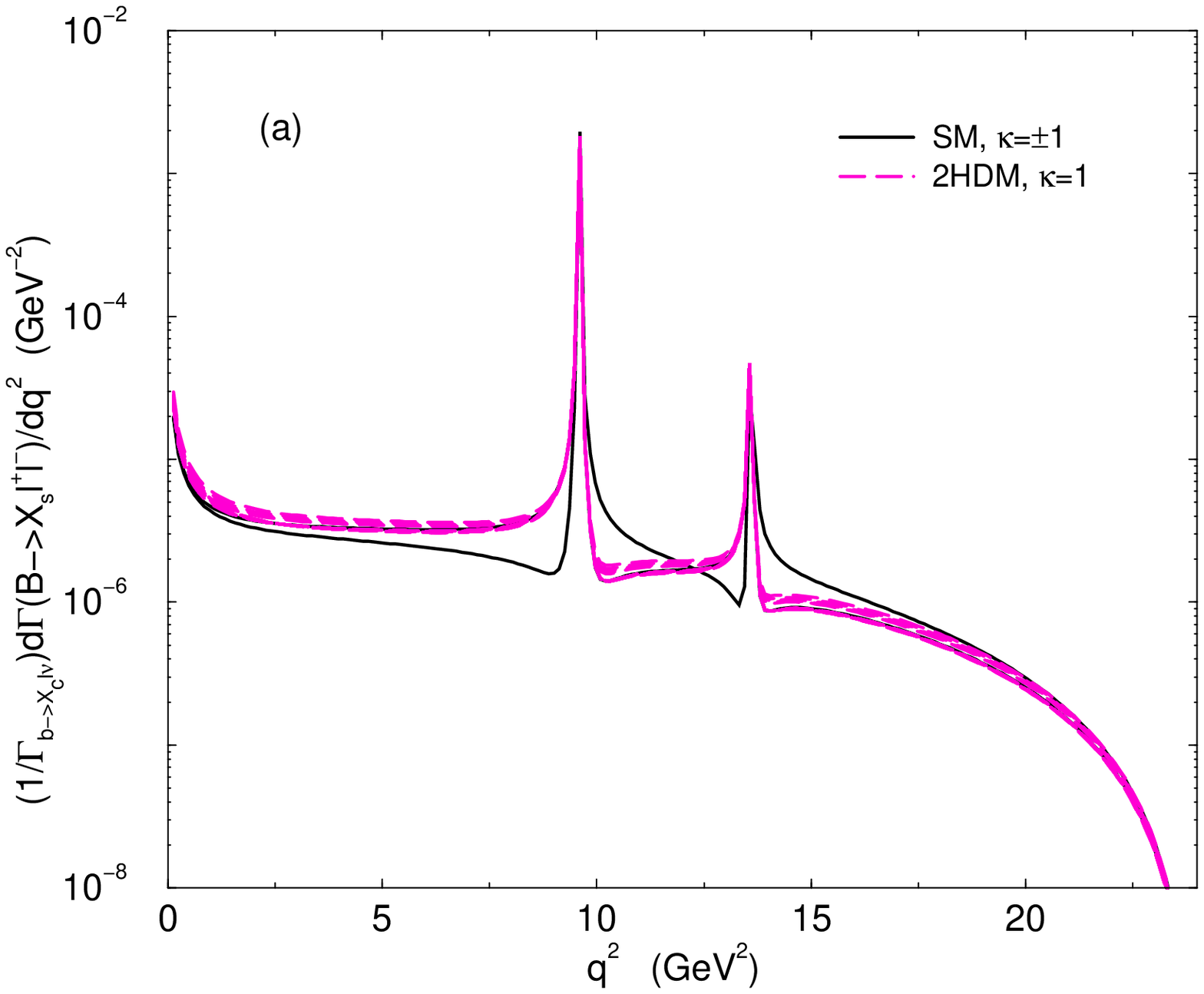}
\includegraphics[width=3.5in]{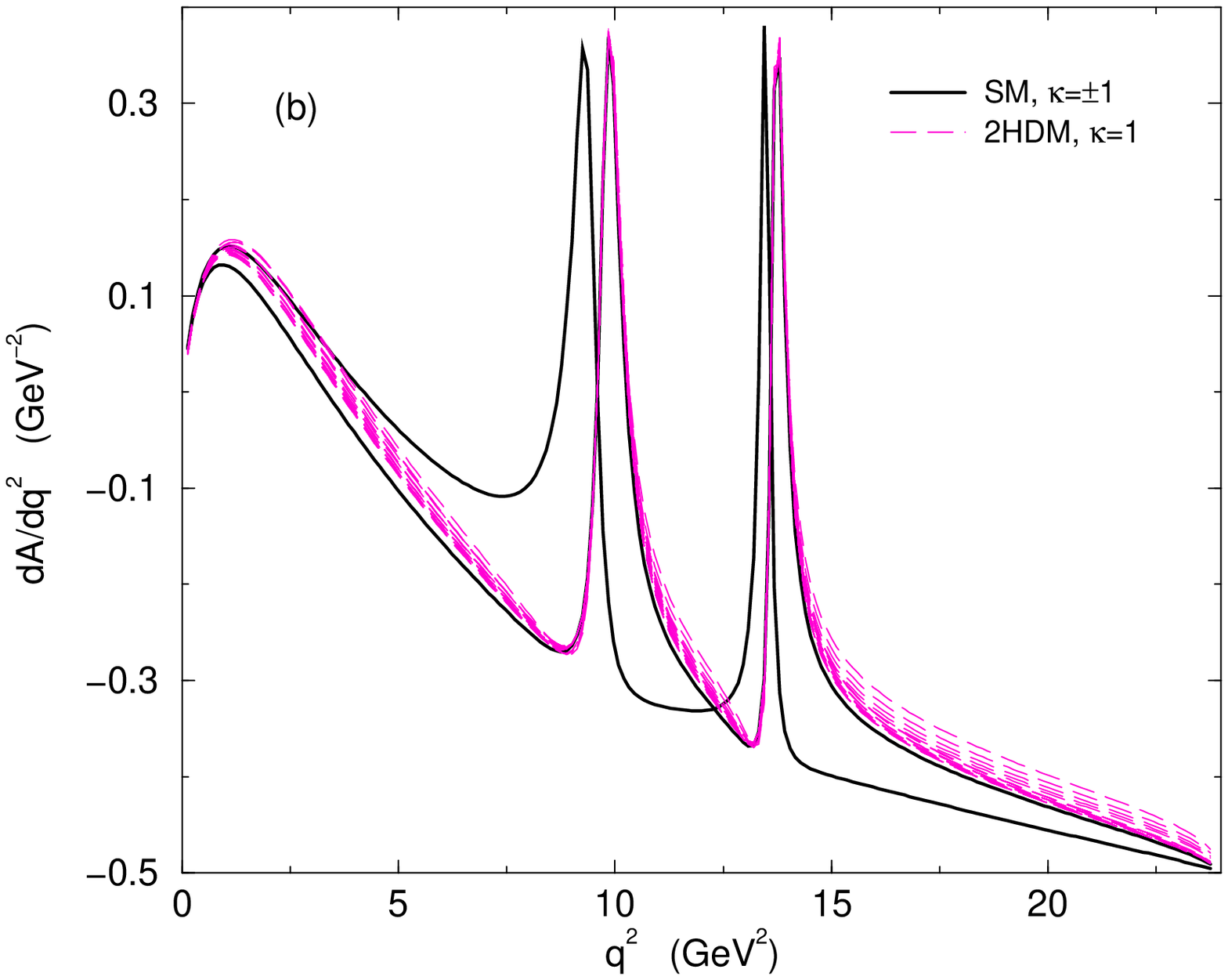}
\end{center}
\caption{
\label{fig3}
(a) Invariant mass and (b) forward-backward asymmetry for the 
two-higgs-doublet model II
with $m_{H^\pm}=400-800$ GeV with an increment of 100 GeV and 
$\tan=1-40$.  The two curves of the SM with $\kappa=\pm1$ are also shown.
}
\end{figure}

We are now ready to investigate the effects of extra charged Higgs 
contributions to $C_i(M_W)$ and in turns to the spectra of leptonic 
invariant mass and forward-backward asymmetry.  Since 2HDMII always pushes
$C_7(M_W)$ more negative, the absolute value of $C_7^{\rm eff}$ increases
and so does the rate of $b\to s\gamma$.  
Using the experimental rate from CLEO: $1\times 10^{-4}< 
BR(b\to s\gamma) < 4.2\times 10^{-4}$ at 95\%CL level \cite{cleo}, 
we limit the range of charged Higgs mass to be $m_{H^\pm} \agt 400$ GeV 
for all $\tan\beta>1$.  
In Fig. \ref{fig3}, we show the invariant mass and forward-backward 
asymmetry for the 2HDMII with $m_H=400-800$ GeV in an increment of 100 GeV and 
$\tan\beta=1,10,20,30,$ and 40.
Here we do not include the effects of fermi motion nor the
leptonic smearing.  It is clear that the results implied by various charged
Higgs mass $m_{H^\pm}\agt 400$ GeV cannot be easily 
distinguished from the SM.  

Next, we are going to use
a model-independent method by varying $C_i(M_W),i=7,9,10$, hoping that it
can cover a wide variety of models of new physics.  We look at each of
$C_7(M_W), C_9(M_W)$, and $C_{10}(M_W)$ while keeping the others at the
SM value.  We shall estimate the range beyond which the resulting spectra
are distinguishable from the SM ones.  We do not look at $C_8(M_W)$ because
$C_8^{\rm eff}(\mu)$ do not enter Eqs. (\ref{width}) and (\ref{asy}) 
directly and therefore the limit on the range of $C_8(M_W)$ is rather loose.
We found that changing the values of $C_{7,9,10}(M_W)$ only changes the
normalization of the continuum part of the invariant mass spectrum, which
is not easy to identify given the experimental uncertainties.  Therefore,
we concentrate on the forward-backward asymmetry.  We confirm that the
forward-backward asymmetry is more sensitive than the
invariant mass spectrum to changes in $C_{7,9}(M_W)$.  
However, for $C_{10}(M_W)$ invariant mass
spectrum appears to change more than the forward-backward asymmetry, but 
still only the normalization of the continuum changes.  We shall discuss it
in a moment.

\begin{figure}[th]
\leavevmode
\begin{center}
\includegraphics[width=4.5in]{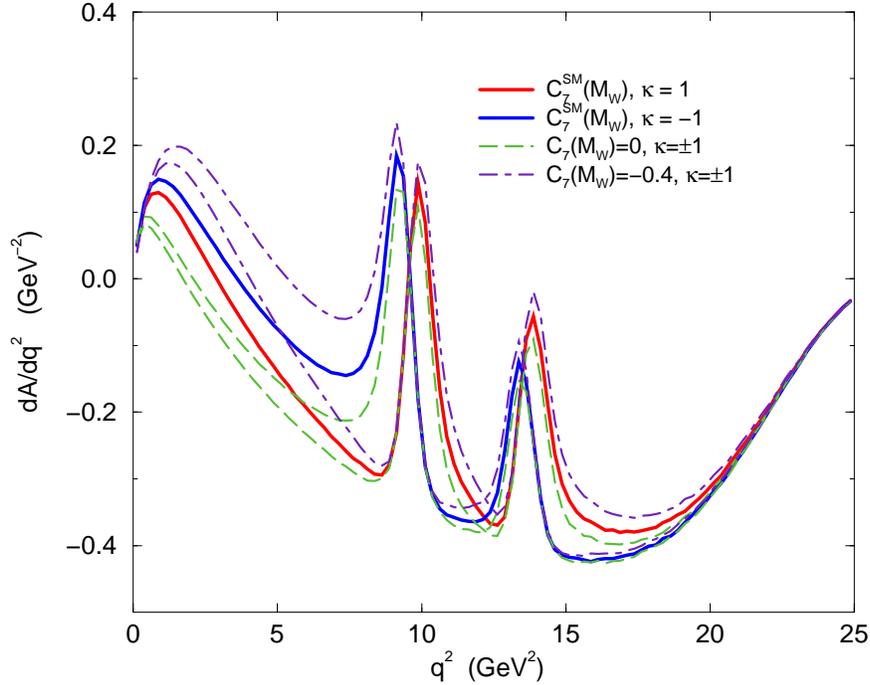}
\end{center}
\caption{
\label{fig4}
The forward-backward asymmetry predicted for $C^{\rm SM}_7(M_W)$, 
$C_7(M_W)=0$, and $C_7(M_W)=-0.4$ with $\kappa=\pm 1$.  The upper curve of 
each set is for $\kappa=-1$ while the lower has $\kappa=1$.  
}
\end{figure}

First we look at $C_7(M_W)$.  We found that the forward-backward asymmetry
is rather sensitive to $C_7(M_W)$ at the small $q^2$ region.
In Fig. ~\ref{fig4}, we show curves for $C^{\rm SM}_7(M_W) \approx -0.2$,
$C_7(M_W)=0$, and $C_7(M_W)=-0.4$ with $\kappa=\pm1$. The region bounded
by the SM curves of $\kappa=\pm1$ shows more or less the uncertainty in
prediction.  We define a $C_7(M_W)$ is distinguishable from the SM 
prediction when it has a significant region not overlapping with the 
SM region.  As seen in Fig. \ref{fig4}, both $C_7(M_W)=0,-0.4$ have a 
region outside the SM region.  
For 
\begin{equation}
\label{c7-2}
C_7(M_W) \alt -0.4\;, \qquad {\rm or} \qquad C_7(M_W) \agt 0 
\end{equation}
the forward-backward asymmetry is further distinguishable from the SM one. 
However, one has to be careful about the range of $C_7(M_W)$ shown in 
Eq. (\ref{c7-2}).
We can apply the SM evolution to evaluate the corresponding range in 
$C_7^{\rm eff}(\mu=m_b)$ and we obtain $C_7^{\rm eff}(m_b) < -0.45$ or
$C_7^{\rm eff}(m_b) > -0.18$, respectively.  The first range 
$C_7^{\rm eff}(m_b) < -0.45$ is already inconsistent with the experimental
rate of $b\to s\gamma$ (the allowed range of $|C_7^{\rm eff}(m_b)|\approx
0.2 - 0.4$.)  The second range, on the other hand, has some overlaps
with the experimentally allowed range.

\begin{figure}[th]
\leavevmode
\begin{center}
\includegraphics[width=4.5in]{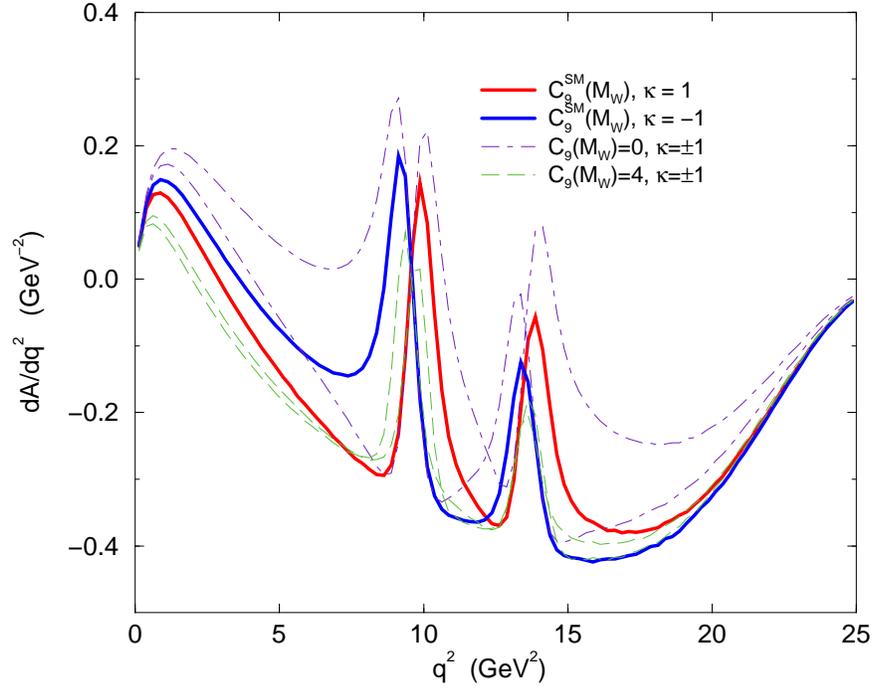}
\end{center}
\caption{
\label{fig5}
The forward-backward asymmetry predicted for $C^{\rm SM}_9(M_W)$, 
$C_9(M_W)=0$, and $C_9(M_W)= 4$ with $\kappa=\pm 1$.  The upper curve of 
each set is for $\kappa=-1$ while the lower has $\kappa=1$.  
}
\end{figure}

\begin{figure}[th]
\leavevmode
\begin{center}
\includegraphics[width=4.5in]{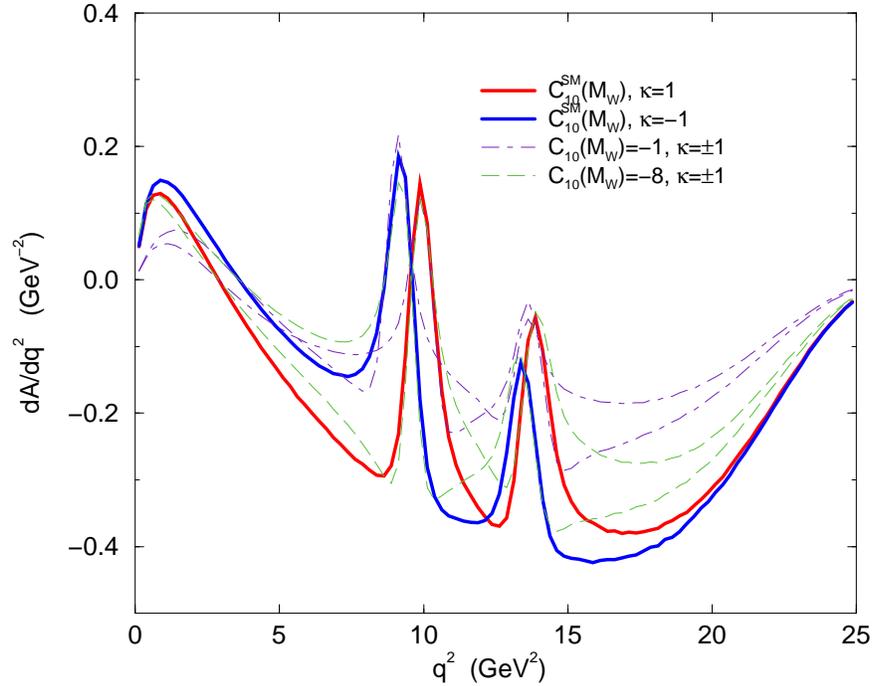}
\end{center}
\caption{
\label{fig6}
The forward-backward asymmetry predicted for $C^{\rm SM}_{10}(M_W)$, 
$C_{10}(M_W)=-1$, and $C_{10}(M_W)= -8$ with $\kappa=\pm 1$.  
The upper curve of each set is for $\kappa=-1$ while the lower has $\kappa=1$.
}
\end{figure}

We now look at $C_9(M_W)$. The SM value for $C_9^{\rm SM}(M_W) \approx 1.6$.
Using the forward-backward asymmetry we found that 
\begin{equation}
C_9(M_W) \alt 0 \;, \qquad {\rm or}\qquad C_9(M_W) \agt 4 \;,
\end{equation}
is needed in order that the resulting spectra is sufficiently different 
from the SM curves, as shown in Fig. \ref{fig5}.
In the SM, $C_{10}(M_W)\approx -4.5$.  Since the forward-backward asymmetry
$dA/dq^2$ is roughly proportional to $C_{10}(M_W)$, as indicated by
the numerator of Eq. (\ref{asy}),
therefore the asymmetry
will not change significantly unless $C_{10}(M_W)$ changes sign.  We 
found that we need a rather large change in $C_{10}(M_W)$ in order for the
forward-backward asymmetry to be distinguishable from the SM curves.
We found, as shown in Fig. \ref{fig6}, 
\begin{equation}
C_{10}(M_W) \alt -8 \;, \qquad {\rm or} \qquad C_{10}(M_W) \agt -1 \;
\end{equation}
is needed.  However, this difference is only marginal and only at the 
large $q^2$ region, where the event rate is relatively low.  

Overall, we have found that we need a rather large change in 
$C_{9,10}(M_W)$ in order to make the forward-backward asymmetry
distinguishable from the SM prediction.  Although $C_7({M_W})$ does not need 
to change a lot for the effect to be seen, the sensitivity range is, however,
severely limited by the experimental rate of $b\to s\gamma$.

\section{Conclusions}

In this report, we have studied the sensitivities of invariant mass and
forward-backward asymmetry of the lepton pair in the decay 
$B\to X_s \ell^+ \ell^-$ to changes in the Wilson coefficients 
$C_{7,9,10}(M_W)$, 
under both the theoretical uncertainties, including the effect of 
$b$-quark fermi motion inside $B$ meson and the unknown phase between the 
perturbative amplitude and the long-distance $\psi$ amplitudes, as well as 
the experimental uncertainty of the measurement of lepton momenta. 
All these uncertainties make the SM prediction become a broad ``region''
that only when new physics predictions go beyond this region can one
say the spectrum is sensitive to new physics.
We found that the sensitivity of the lepton forward-backward asymmetry
is rather weak to changes in $C_{9,10}(M_W)$.  Only when $C_{9,10}(M_W)$
change substantially will the asymmetry be distinguishable from the SM
prediction.  For $C_7(M_W)$ the lepton forward-backward asymmetry is more
sensitive but, however, a large part of the sensitivity range is already
ruled out by the $b\to s \gamma$ rate.

\vspace{0.3in}

This work was supported by the DOE under contracts No. DE-FG03-91ER40674 and 
by the Davis Institute for High Energy Physics.


\begin{thebibliography}{99}

\bibitem{cleo}M.S. Alam {\it et al.} (CLEO Coll.), Phys. Rev. Lett. {\bf 74},
2884 (1995).

\bibitem{back}
A. Ali, T. Mannel, and T. Morozumi, Phys. Lett. {\bf B273}, 505 (1991);
D. Liu, Phys. Rev. {\bf D52}, 5056 (1995);
Melikhov, N. Nikitin, and S. Simula (Rome, ISS), Phys. Lett. {\bf B430}, 332
(1998).

\bibitem{buras}G. Buchalla, A. Buras, and M. Lautenbacher, 
Rev. Mod. Phys. {\bf 68}, 1125 (1996).

\bibitem{psi}N. Deshpande, J. Trampeti\'{c}, and K. Panose, Phys. Rev. 
{\bf D39}, 1461 (1989);C. Lim, T. Morozumi, and A. Sanda, Phys. Lett. 
{\bf B218}, 343 (1989).

\bibitem{donnell}P. O'Donnell and H. Tung, Phys. Rev. {\bf D43}, 2067 (1991).

\bibitem{altarelli}G. Altarelli {\it et al.}, Nucl. Phys. {\bf B208}, 365
(1982).

\bibitem{wise}B. Grinstein, R. Springer, and M. Wise, 
Nucl. Phys. {\bf B339}, 269 (1990).

\bibitem{ali}A. Ali {\it et al.}, Phys. Rev. {\bf D55}, 4105 (1997);
A. Ali and G. Hiller, hep-ph/9807418.


\bibitem{ligeti}Z. Ligeti and M. Wise, Phys. Rev. {\bf D53}, 4937 (1996).

\bibitem{hq}J. Chay, H. Georgi, and B. Grinstein, Phys. Lett. {\bf B247}, 
399 (1990).

\bibitem{cleo-psi}CLEO Collaboration, R. Balest {\it et al.}, Phys. Rev. 
{\bf D52}, 2661 (1995).

\bibitem{hewett}J. Hewett and  J. Wells, Phys. Rev. {\bf D55}, 5549 (1997).

\end{thebibliography}
\end{document}